\pdfoutput=1
\documentclass[
 reprint, amsmath, amssymb,
 aps, floatfix,
 superscriptaddress,
 prl
 ]{revtex4-2}
\usepackage{comment}
\usepackage[mathlines]{lineno}
\usepackage{color}
\usepackage{hyperref}
\usepackage{color}
\usepackage{graphicx}
\usepackage{dcolumn}
\usepackage{bm}
\usepackage{here}
\usepackage{braket}
\usepackage{multirow}
\usepackage{pgfplots}
\usetikzlibrary{backgrounds}
\pgfplotsset{compat=1.16}
\usepgfplotslibrary{fillbetween}

\DeclareUnicodeCharacter{2212}{-}

\begin{document}

\title{Fixed-point tensor is a four-point function}

\author{Atsushi Ueda}
\email{aueda@issp.u-tokyo.ac.jp}
\affiliation{%
 Institute for Solid State Physics, University of Tokyo, Kashiwa 277-8581, Japan
}
\author{Masahito Yamazaki}
\email{masahio.yamazaki@ipmu.jp}
\affiliation{Kavli IPMU (WPI), UTIAS, The University of Tokyo, Kashiwa, Chiba 277-8583, Japan}
\affiliation{Center for Data-Driven Discovery, Kavli IPMU (WPI), UTIAS, The University of Tokyo, Kashiwa, Chiba 277-8583, Japan}
\affiliation{Trans-Scale Quantum Science Institute, The University of Tokyo, Tokyo 113-0033, Japan}

\date{\today}

\begin{abstract} 
Through coarse-graining, tensor network representations of a two-dimensional critical lattice model flow to a universal four-leg tensor, corresponding to a conformal field theory (CFT) fixed-point. We computed explicit elements of the critical fixed-point tensor, which we identify as the CFT four-point function. This allows us to directly extract the operator product expansion coefficients of the CFT from these tensor elements. Combined with the scaling dimensions obtained from the transfer matrix, we determine the complete set of the CFT data from the fixed-point tensor for any critical unitary lattice model.
\end{abstract}


\maketitle

{\it{Introduction.}--- }Renormalization group (RG) \cite{Stuckelberg:1951,*Stuckelberg:1953,Gell-Mann:1954yli,Bogolubov:1955-1,*Bogolubov:1955-2} is one of the most profound concepts in contemporary physics.
RG theory has significantly deepened our understanding of the universality of critical phenomena \cite{Wilson:1973jj,RevModPhys.49.267}. We now understand that each universality class is described by an RG fixed-point (FP) theory under the RG transformation, which theory can be represented \cite{Polchinski:1987dy,Luty:2012ww} as a conformal field theory (CFT) \cite{Belavin:1984vu}. 
Universal behavior, such as critical exponents, can then be elucidated from the 
CFT data, which include central charges, scaling dimensions, and operator product expansion (OPE) coefficients \cite{cardy1996scaling,DiFrancesco:1997nk,zinn2021quantum}. It is therefore of paramount importance to identify this CFT data for a given ultraviolet (UV) theory (such as a lattice model).~\footnote{The conformal bootstrap \cite{Ferrara:1973yt,Polyakov:1974gs,Rattazzi:2008pe,RevModPhys.91.015002} is a notable technique that has successfully computed the CFT data for 
e.g.\ the three-dimensional Ising model to a high precision \cite{El-Showk:2012cjh}. 
Despite its successes, the main focus of the conformal bootstrap is to constrain the possible 
parameter spaces of CFTs, and it is often not sufficient if one wishes to calculate the CFT data for a specific lattice model. In this sense, our results complement the bootstrap program.}.

While the analysis of the real-space RG transformation has a long history \cite{PhysicsPhysiqueFizika.2.263}, tensor network renormalization (TNR)~\cite{PhysRevLett.99.120601,PhysRevLett.115.180405,PhysRevB.95.045117,PhysRevLett.118.110504,PhysRevLett.118.250602,PhysRevB.97.045111,homma2023nuclear}
has recently emerged as a reliable numerical implementation of the real-space RG. The application of TNR has demonstrated that the tensor-network representation of the Boltzmann weights converges to a FP tensor, representing the RG fixed point. 

There are several motivations for studying the FP tensors. 

First, we expect that the FP tensor encodes the CFT data of the FP theory. Gu and Wen have established a method for calculating the central charge and scaling dimensions for fixed-point tensors, a procedure that has since become standard~\cite{PhysRevB.80.155131}. It remains an intricate and challenging problem, however, to compute the OPE coefficients of the FP CFT~\cite{PhysRevLett.116.040401,PhysRevResearch.4.023159,PhysRevB.108.024413,guo2023tensor}.

Second, determination of the fixed-point tensor can 
facilitate concrete realizations of the RG flow. Recently, Kennedy and Rychkov initiated a rigorous study of the RG using tensor networks~\cite{Kennedy_2022,kennedy2023tensor}. Employing simple low-temperature and high-temperature fixed-point tensors, they successfully demonstrated the stability of the corresponding fixed points. Nevertheless, the application of similar arguments to critical fixed points remains unachieved, given that even their tensor network representations are not fully understood.

Third, precise expressions of the fixed-point tensors will serve as a robust benchmark for evaluating the precision of different tensor-network algorithms. 
A number of algorithms boasting increased accuracy have been developed to determine the FP tensor, but 
there remain uncertainties in 
selecting the superior option due to our limited understanding of the exact expression of the fixed-point tensor. 

In this Letter, we introduce an exact tensor network representation of critical RG fixed points,
thereby solving the problem of numerically determining the full defining data of the FP CFT. 
We anticipate that our findings will serve as a pivotal contribution 
in practical computations of the FP theory on the one hand, and 
towards the rigorous substantiation of RG theory, on the other.

{\it{Fixed-point tensor.}--- }To simulate two-dimensional statistical models, we use the tensor network methods, where the local Boltzmann weight is represented as a four-legged tensor $T^{(0)}$. We obtain the transfer matrix in the $y$-direction if we contract $L$ copies of the four-leg tensors along a circle in the $x$-direction; we obtain the partition function $Z(L, T^{(0)})$ if we contract $L\times L$ copies along the torus in the $x, y$-directions. We can also contract $L\times L$ copies of $T^{(0)}$ in the $x, y$-directions, but with endpoints un-contracted (as in the right-hand side of the Figure below).
In the limit $L\to \infty$,
this contracted tensor converges to a  universal rank-four tensor $T^{*}$ with an infinite bond dimension that corresponds to the fixed-point of the RG transformation:
\begin{equation}
    \includegraphics[width=86mm]{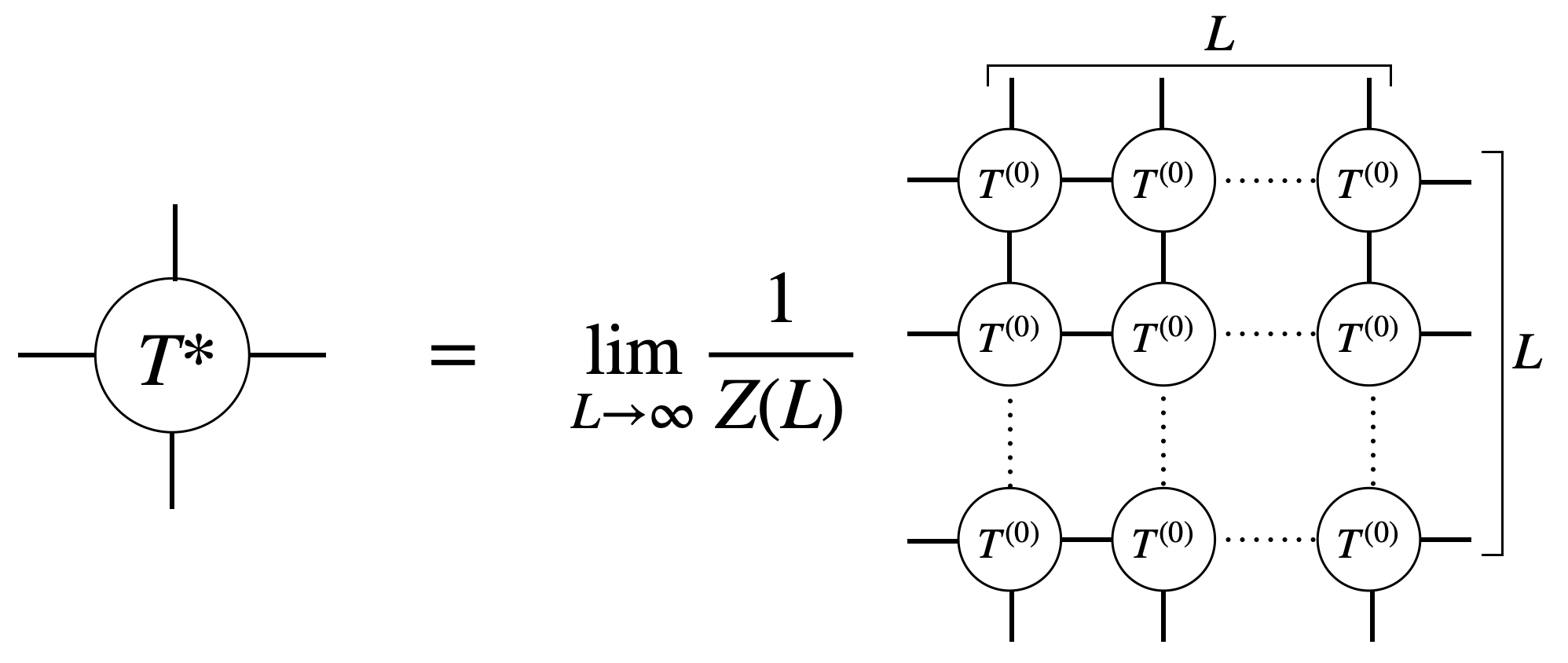}\nonumber
\end{equation}
This tensor $T^*$ is called the FP tensor.

If the original tensor $T^{(0)}$ has $\mathrm{D}_4$ symmetry, $T^{*}$ also respects it. This allows the decomposition of the FP tensor into a pair of two identical three-leg tensors $S^*$:
\begin{align}
    \begin{matrix}
    \includegraphics[width=77mm]{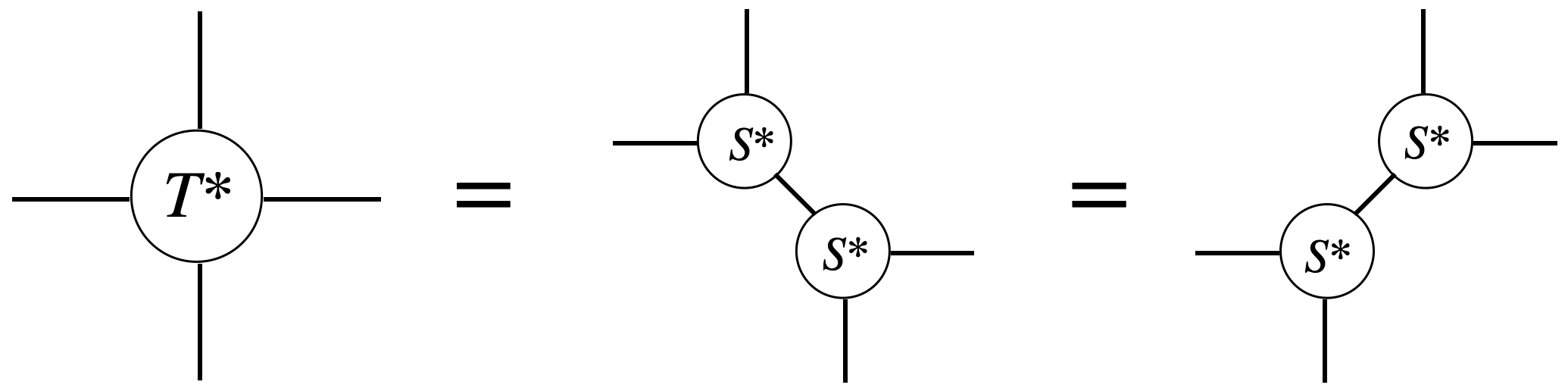}
    \end{matrix}
    \label{T_in_S}
\end{align}

The FP tensor $T^{*}$ has gauge degrees of freedom that change the basis of each leg. The insertion of the gauge transformation (unitary operators) does not change the spectral property of the FP tensor. In the following, we fix the gauge so that each index of the FP tensor is labeled by the eigenstates of the Hamiltonian $L_0+\bar{L}_0$ on a cylinder,
where $L_n$ ($\bar{L}_n$) are the standard generators of the left-moving (right-moving) Virasoro algebras.
By the state-operator correspondence, we can label these states by a set of operators $\phi_\alpha$,
among which we will find the identity operator $\phi_{1}$ with the lowest scaling dimension. \footnote{Note that the label $\alpha$ refers to both the primaries and the descendants of the Virasoro algebra.} In tensor-network representations, the projector to this basis can be found by diagonalizing the transfer matrix as follows \cite{PhysRevB.80.155131}: 
\begin{align}
    \begin{matrix}\includegraphics[width=77mm]{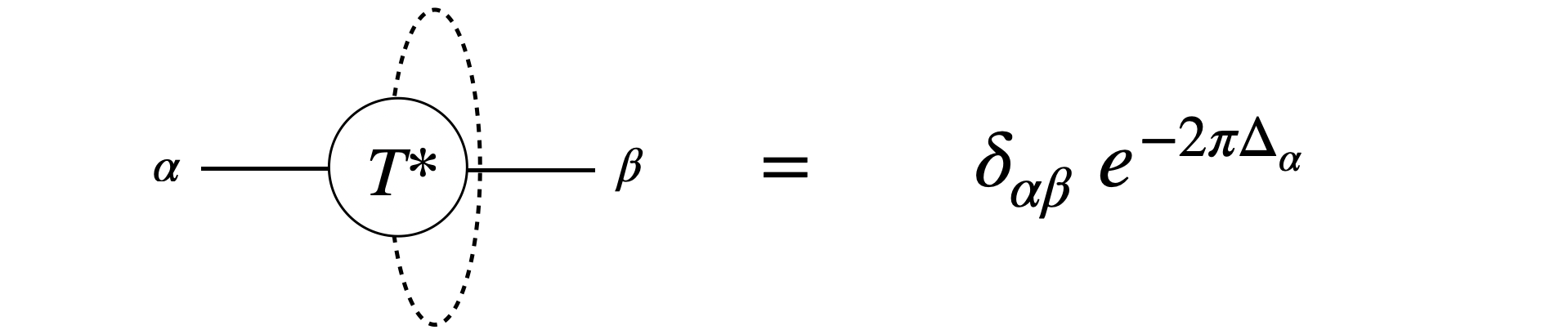} 
    \end{matrix}
    \label{x_two_point}
\end{align}
\begin{figure}[tb]
    \centering
    \includegraphics[width=80mm]{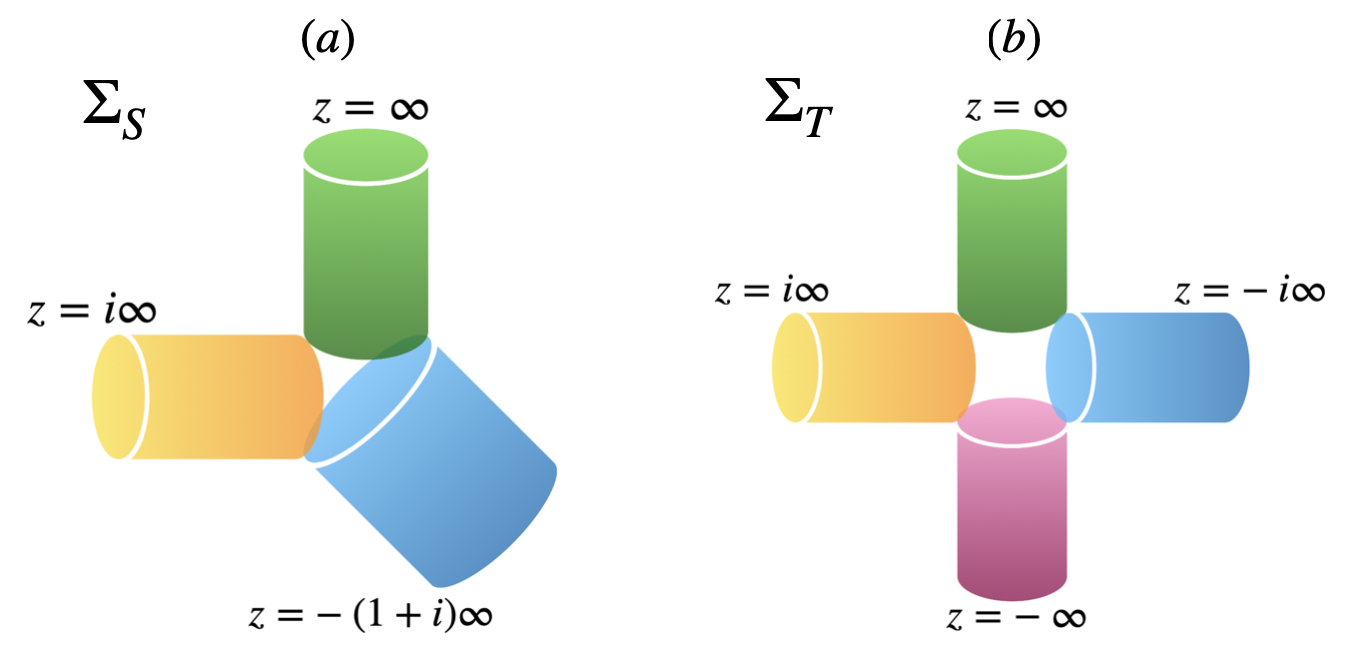}
    \caption{$(a)$ The path-integral representation of the tensor $S^*_{\alpha\beta\gamma}$. $(b)$ The path-integral representation of the tensor $T^*_{\alpha\beta\gamma\delta}$.}
    \label{Tensor_S}
\end{figure}
In the following, we choose the states $\alpha, \beta, \dots$ to be primary operators.
\par{\it{Main Results.}--- }Let us now state the main results of this paper.
First, the three-leg tensor $S^*$
 is proportional to the three-point functions of the FP CFT on the complex plane:
\begin{align}
    \frac{S^*_{\alpha\beta\gamma}}{S^*_{111}}&= \langle\phi_\alpha(-x_S)\phi_\beta(ix_S)\phi_\gamma(0)\rangle_{\rm pl} .
    \label{x_three_point}
\end{align}
Second, the four-leg FP tensor 
 determines the four-point functions of the FP CFT as 
\begin{align}
    \frac{T^*_{\alpha\beta\gamma\delta}}{T^*_{1111}} &=\langle\phi_\alpha(-x_T)\phi_\beta(ix_T)\phi_\gamma(x_T)\phi_\delta(-ix_T)\rangle_{\rm pl}.
    \label{x_four_point}
\end{align}
These equalities hold when we choose the values $x_S=e^{\pi/4}$ and $x_T=e^{\pi/2}/{2}$. 

We can now reproduce the {\it full} defining data for the FP CFT.
Recall that we can extract the scaling dimensions $\Delta_{\alpha}$ operators from Eq.~\eqref{x_two_point}.
The remaining data is the OPE coefficients $C_{\alpha\beta\gamma}$ of the operators $\phi_\alpha$, which can be extracted by 
applying a conformal transformation to Eq.~\eqref{x_three_point}:
\begin{align}
\frac{S^*_{\alpha\beta\gamma}}{S^*_{111}}&=\frac{C_{\alpha\beta\gamma}}{x_S^{\Delta_\beta+\Delta_\gamma-\Delta_\alpha}x_S^{\Delta_\gamma+\Delta_\alpha-\Delta_\beta}(\sqrt{2}x_S)^{\Delta_\alpha+\Delta_\beta-\Delta_\gamma}},\nonumber\\
&=\frac{2^{\Delta_\gamma}C_{\alpha\beta\gamma}}{(\sqrt{2}x_S)^{\Delta_\alpha+\Delta_\beta+\Delta_\gamma}} .
    \label{S_in_C}
\end{align}

Equation \eqref{T_in_S} represents
the equivalence of two different decompositions ($s$- and $t$-channels) of the four-point function into a pair of three-point functions, i.e.\ the celebrated crossing relation of the CFT.

To better understand Eqs.~(\ref{x_three_point}-\ref{x_four_point}),
we apply conformal transformations to the two equations to obtain 
\begin{align}
    \frac{S^*_{\alpha\beta\gamma}}{S^*_{111}}
    &= e^{-\frac{\pi}{4}(\Delta_\alpha+\Delta_\beta+\Delta_\gamma)}\langle\phi_\alpha(-1)\phi_\beta(i)\phi_\gamma(0)\rangle_{\rm pl}\label{S_three_point}, \\
    \frac{T^*_{\alpha\beta\gamma\delta}}{T^*_{1111}}
    &= \left(\frac{e^{\frac{\pi}{2}}}{2}\right)^{-\Delta_{\rm tot}}\langle\phi_\alpha(-1)\phi_\beta(i)\phi_\gamma(1)\phi_\delta(-i)\rangle_{\rm pl}\label{four_point},
\end{align}
where $\Delta_{\rm tot} \equiv \Delta_\alpha+\Delta_\beta+\Delta_\gamma+\Delta_\delta$. 

Equations~(\ref{S_three_point}-\ref{four_point}) naturally arise from conformal mappings~\cite{PhysRevB.107.155124,PhysRevB.105.125125}. Once we fix the basis for the fixed-point (FP) tensor, each index corresponds to the states of CFT. Utilizing state-operator correspondence, the normalized wave function of the first index of $S^*$, for instance, is created by inserting $\phi_\alpha$ in the future infinity of the cylinder as follows:
$$|\phi^1\rangle=\left(\frac{2\pi}{L}\right)^{-\Delta_\alpha}\lim_{z\rightarrow\infty}e^{2\pi z\Delta_\alpha/L}\phi_\alpha(\infty)|I^{\rm cyl}\rangle,$$
where $|I^{\rm cyl}\rangle$ represents the ground state corresponding to the identity operator.
Subsequently, the FP tensors $S^*$ and $T^*$ can be expressed by the path integral on the manifolds $\Sigma_S$ and $\Sigma_T$, respectively, as illustrated in Fig.~\ref{Tensor_S}. Then, the FP-tensor elements are
\begin{align}
    \frac{S^*_{\alpha\beta\gamma}}{S^*_{111}}&= \langle\phi_\alpha(\infty)\phi_\beta(i\infty)\phi_\gamma(-(1+i)\infty)\rangle_{\Sigma_S},\\
    \frac{T^*_{\alpha\beta\gamma\delta}}{T^*_{1111}} &=\langle\phi_\alpha(-\infty)\phi_\beta(i\infty)\phi_\gamma(\infty)\phi_\delta(-i\infty)\rangle_{\Sigma_T}.
\end{align}
$\Sigma_S$ and $\Sigma_T$ can be mapped the complex plane $w$ by using (cf.\ \cite{Mandelstam:1973jk}),
\begin{align}
    z_S &= \frac{L}{2\pi}[-\ln(w-i)-i\ln(w+1)+(1+i)\ln w]\label{confmap_S},\\
    z_T &= \frac{L}{2\pi}\left[\ln\left(\frac{w+i}{w-i}\right)+i\ln\left(\frac{w-1}{w+1}\right)\right].
\end{align}
Each operator in the $z$-coordinate transforms accordingly as 
\begin{align}
\frac{S^*_{\alpha\beta\gamma}}{S^*_{111}}&=\langle\phi_\alpha(-1)\phi_\beta(i)\phi_\gamma(0)\rangle_{\rm pl}\prod_{n\in(\alpha,\beta,\gamma)}|J_n|^{\Delta_n},\nonumber\\
    \frac{T^*_{\alpha\beta\gamma\delta}}{T^*_{1111}}&=\langle\phi_\alpha(-1)\phi_\beta(i)\phi_\gamma(1)\phi_\delta(-i)\rangle_{\rm pl}\prod_{n\in(\alpha,\beta,\gamma,\delta)}|J_n|^{\Delta_n}\nonumber,
\end{align}
where $|J_n| = |\left(\frac{2\pi}{L}\right)^{-1}\lim_{z\rightarrow\zeta\infty}e^{2\pi z\zeta^*/(L|\zeta|)}|w'(z)|$, and $\zeta\infty$ is the coordinate of the index in the originate manifold. The resulting $|J_n|$ are $e^{-\pi/4}$ and $2e^{-\pi/2}$, respectively, being consistent with Eqs.~(\ref{S_three_point}-\ref{four_point}). Detailed calculations are presented in the supplemental material.

\begin{figure}[tb]
    \centering
    \includegraphics[width=80mm]{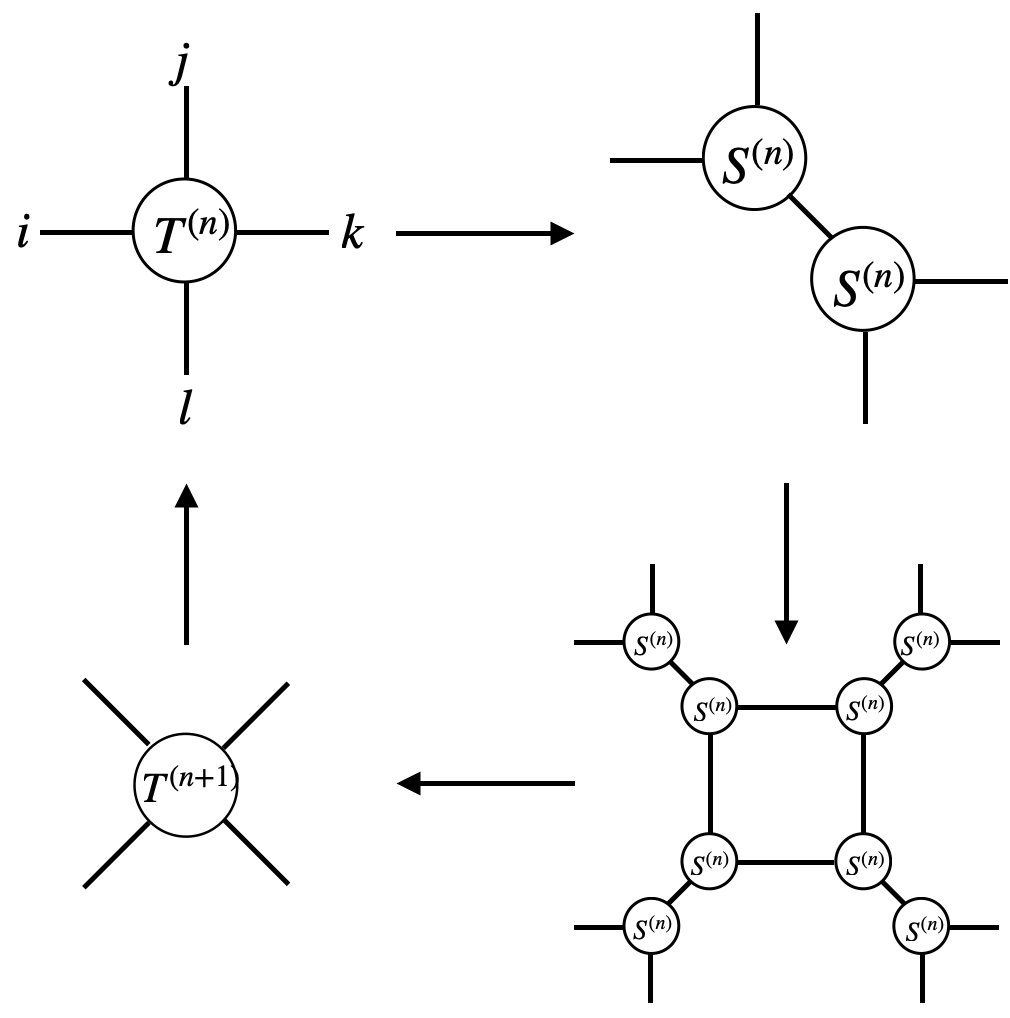}
    \caption{The pictorial description of the tensor renormalization group. The decomposition is done so that $T^{(n)}$ is a good approximation of the local Boltzmann weights of $L=\sqrt{2}^n$. In the tensor network renormalization(TNR) scheme, filtering of local entanglement is introduced.}
    \label{trg}
\end{figure}

{\it{Numerical fixed point tensor.}--- }Let us provide numerical confirmations of our main results using tensor renormalization group (TRG)~\cite{PhysRevLett.99.120601}.
TRG is a numerical technique devised to calculate effective $L \times L$ tensor networks. In our study, our interest lies in computing those of large system sizes to obtain a tensor that is as close as possible to the FP tensor. However, performing an exact contraction is exponentially difficult, prompting us to focus on extracting low-lying spectral properties. TRG seeks to circumvent this issue by employing the principles of the renormalization group theory. Each coarse-graining step entails decompositions and recombinations as depicted in Fig.~\ref{trg}. Truncation, parameterized by the bond dimension $D$, is performed to maintain the tractability of numerical computation. However, it is important to note that this scheme is considered {\it exact} when $D = \infty$, and thus, employing larger $D$ improves the numerical accuracy. Additionally, we impose ${\rm D_4}$ symmetry in TRG. The details can be found in the supplemental material.
\begin{figure}[tb]
    \centering
    \includegraphics[width=86mm]{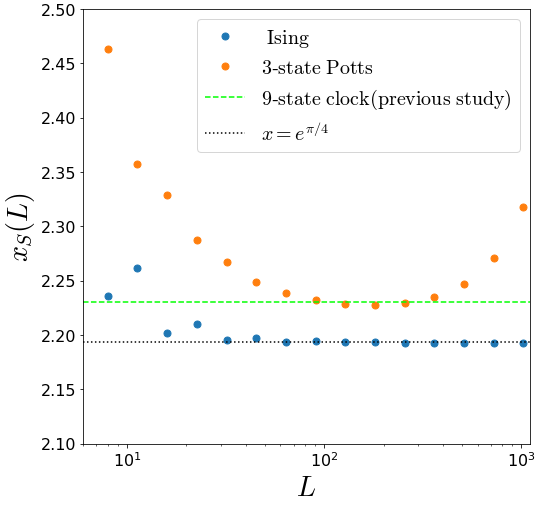}
    \caption{Estimation of $x_S(L)$ from TRG at $D=96$. The values of $x(L)$ from both the Ising and 3-state Potts model converge to the theoretical value $x_S=e^{\pi/4}$ denoted by a black dotted line. We plot $x_S=2.23035$ obtained from Loop-TNR~\cite{PhysRevLett.118.110504} on the critical 9-state clock model~\cite{PhysRevResearch.4.023159} with a lime dashed line. The 3-state Potts model exhibits a deviation for $L>100$ because simulating systems with higher central charges involves larger numerical errors.}
    \label{x_fs}
\end{figure}

{\it{Tests on critical lattice models.}---}Let us first test the value $x_S=e^{\pi/4}$ in Eq.~\eqref{S_three_point}, by computing  $x_S$ from the critical Ising and 3-state Potts models. 
Given Eq.~\eqref{S_three_point}, we can numerically compute the OPE coefficients $C_{\alpha\beta\gamma}$ from Eq.~\eqref{S_in_C}. 
We define $x_S(L)$ by solving Eq.~\eqref{S_in_C} to be 
\begin{align}
    x_S(L) \equiv \frac{1}{\sqrt{2}}\left(\frac{2^{\Delta_\gamma}C_{\alpha\beta\gamma}}{S_{\alpha\beta\gamma}(L)}\right)^{1/(\Delta_\alpha+\Delta_\beta+\Delta_\gamma)}.
\end{align}
Each model has a primary operator $\epsilon$, called the energy and the thermal operator, respectively. Since $C_{\epsilon\epsilon1}=1$, $x_S(L)$ can be computed from the finite-size three-leg tensor $S_{\epsilon\epsilon1}(L)$.

Figure~\ref{x_fs} shows the value of $x_S(L)$ obtained from TRG at the bond dimension $D=96$. The numerically-derived $x_S(L)$'s for both models converge to the theoretical value of $e^{\pi/4}$. The noticeable increase in amplitude for the 3-state Potts model at $L>10^2$ is attributed to the effect of the finite bond dimension. It is worth noting that our value for $x_S$ deviates slightly from the value $x_S=2.23035$ from a previous study on the 9-state clock model~\cite{PhysRevResearch.4.023159}. We speculate that this minor deviation is due to the finite bond-dimension effect because higher central charges lead to more pronounced numerical errors~\cite{PhysRevB.108.024413}. For the system size $L=2048$ and bond dimension $D=96$, we ascertain $x_S = 2.193257$ for the Ising model, a value remarkably close to $e^{\pi/4} = 2.193280$.

\begin{figure}[tb]
    \centering
    \includegraphics[width=86mm]{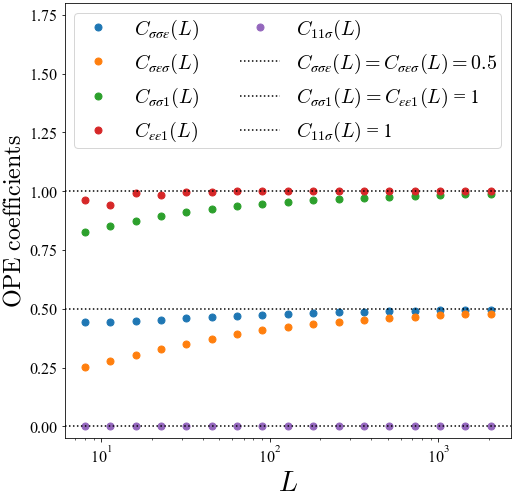}
    \caption{The OPE coefficients evaluated by setting $x_S = e^{{\pi/ 4}}$. The black dotted lines denote the theoretical values 0.5 and 1. }
    \label{ope_finitesize}
\end{figure}

Once we are certain of the value $x_S=e^{\pi/4}$, we can 
verify Eq.~\eqref{S_three_point} for all the OPE coefficients, which are computed from the three-leg tensor $S$ as 
\begin{align}
    C_{\alpha\beta\gamma}(L) = (\sqrt{2}\, e^{\pi/4})^{\Delta_\alpha+\Delta_\beta+\Delta_\gamma}2^{-\Delta_\gamma}S_{\alpha\beta\gamma}(L).\label{ope_fomula}
\end{align}
The results are exhibited in Fig.~\ref{ope_finitesize}. The finite-size effect originates from the twist operator at the branch points~\cite{PhysRevB.107.155124,PhysRevB.105.125125}, whose scaling is universal. The detailed analysis is discussed in the supplemental material.\\

\begin{figure}
    \centering
    \includegraphics[width=86mm]{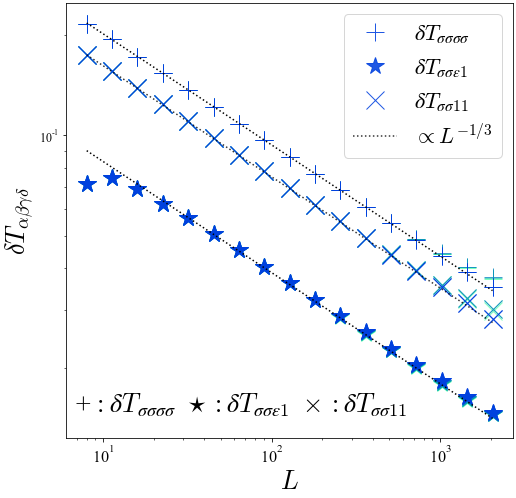}
    \caption{The finite-size effect of the fixed point tensor $\delta T_{\alpha\beta\gamma\delta} \equiv \langle\phi_{\alpha}\phi_j{\beta}\phi_{\gamma}\phi_{\delta}\rangle-T_{\alpha\beta\gamma\delta}(L)$ for various bond dimensions $D=84,\ 88,\ 92,$ and 96. We plot $\delta T_{\alpha\beta\gamma\delta}$ of $\sigma\sigma\sigma\sigma$(``$+$"), $\sigma\sigma\epsilon1$(``$\star$"), and $\sigma\sigma11$(``$\times$") with darker colors for higher bond dimensions. The difference converges to zero for $L\rightarrow\infty$ with the power-law $\sim L^{-1/3}$.}
    \label{4-point_fs}
\end{figure}

We next computed four-point tensors 
$T_{\alpha \beta \gamma \delta}$ and compared with the theoretical values from Eq.~\eqref{four_point},
where the explicit forms of the four-point functions are listed in the supplemental material.
The result is consistent up to two digits for most tensor elements, as shown in Table~\ref{fptensor_table}. The exceptions are $T_{\sigma\sigma\sigma\sigma}$ and $T_{\sigma\sigma11}$, whose numerical values deviate approximately 5\% from the theoretical values. As for $T_{\sigma\sigma\epsilon1}$, the deviation is almost 24\%.
This discrepancy, however, can be attributed to finite-size effects and becomes negligible for infinite system sizes. To illustrate this, we define the finite-size deviation as
\begin{align}
    \delta T_{\alpha\beta\gamma\delta}\equiv T^*_{\alpha\beta\gamma\delta}-T_{\alpha\beta\gamma\delta}(L).\nonumber
\end{align}
Figure~\ref{4-point_fs} presents the values of $\delta T_{\sigma\sigma\sigma\sigma}(L)$, $\delta T_{\sigma\sigma\epsilon 1}(L)$, and $\delta T_{\sigma\sigma11}(L)$ obtained from TRG calculations. A clear power-law decay with respect to the system size is observed, supporting the claim that the 
large deviations for those elements are finite-size effects. However, it is worth mentioning that the exponent closely approximates $\sim L^{-1/3}$, hinting at the existence of an underlying theory that might account for this.
\begin{table}[tb]
\begin{ruledtabular}
\caption{The comparison of the numerically-obtained fixed-point tensor $T_{\alpha \beta \gamma \delta}$ at $L=2048$ and the exact four-point function $\langle\phi_{\alpha}(-x_T)\phi_{\beta}(ix_T)\phi_{\gamma}(x_T)\phi_{\delta}(-ix_T)\rangle_{\rm pl}$ of the Ising model with $x_T = e^{\pi/2}/2$. \label{fptensor_table} }
\begin{tabular}{lllll}
                                   & $T_{\alpha \beta \gamma \delta}\, (L=2048)$ & $\langle\phi_{\alpha}\phi_{\beta}\phi_{\gamma}\phi_{\delta}\rangle$ \\
                                   \hline
$1111$                             & 1          & 1                                        \\
$\sigma\sigma\sigma\sigma$         & 0.610       & 0.645                                    \\
$\sigma\sigma\epsilon\epsilon$     & 0.0714      & 0.0716                                    \\
$\sigma\epsilon\sigma\epsilon$     & 0.000      & 0                                    \\
$\epsilon\epsilon\epsilon\epsilon$ & 0.0168      & 0.0168 \\
$\sigma\sigma\epsilon1$         & 0.0618       & 0.0765                                    \\
$\sigma\epsilon\sigma1$         & 0.133       & 0.140                                    \\
$\sigma\sigma\sigma1$         & 0.000       & 0                                    \\
$\epsilon\epsilon\epsilon1$         & 0.001       & 0                                    \\
$\sigma\sigma11$         & 0.708       & 0.736                                    \\
$\sigma1\sigma1$         & 0.639       & 0.675                                    \\
$\epsilon\epsilon11$         & 0.0863       & 0.0864                                    \\
$\epsilon1\epsilon1$         & 0.0439       & 0.0432                                    \\
$\epsilon\sigma11$         & 0.000       & 0                                  \\
\end{tabular}
\end{ruledtabular}
\end{table}

{\it{Acknowledgement}---} We would like to thank Jacob Bridgeman, Clement Delcamp, Jutho Haegeman, Rui-Zhen Huang, Kansei Inamura, Andreas L\"{a}uchli, Laurens Lootens, Masaki Oshikawa, Slava Rychkov, Luca Tagliacozzo, Frank Verstraete and Yunqin Zheng for helpful discussions. A.~U.~is supported by the MERIT-WINGS Program at the University of Tokyo, the JSPS fellowship (DC1). He was supported in part by MEXT/JSPS KAKENHI Grants No.\ JP21J2052. M.~Y.~was supported in part by the JSPS Grant-in-Aid for Scientific Research (19H00689, 19K03820, 20H05860, 23H01168), and by JST, Japan (PRESTO Grant No.\ JPMJPR225A, Moonshot R\&D Grant No.\ JPMJMS2061).

{\it{Source Availability.}--} Our numerical data and analysis codes for the Ising fixed-point are publicly available at  \url{https://github.com/dartsushi/TRG_D4_symmetry}.

\bibliographystyle{apsrev4-2}
\bibliography{apssamp}

\newpage
\section{Supplemental material}
\subsection{Conformal mapping of S}

The three-leg tensor $S^*_{\alpha\beta\gamma}$ represents the three-sided thermofield double state corresponding to the geometry in Fig.~\ref{Tensor_S}(a). This manifold $\Sigma_S$ is mapped to the plane by a conformal mapping 
\begin{align}
    z = \frac{L}{2\pi}[-\ln(w-i)-i\ln(w+1)+(1+i)\ln w],
    \label{z_to_w}
\end{align}
which maps 
the three points in $\Sigma_S$, $(z_1,z_2,z_3) = (\infty,i\infty,-(1+i)\infty)$, to $(w_1,w_2,w_3)=(i,-1,0)$. 
Then, the tensor element is
\begin{align}
\frac{S^*_{\alpha\beta\gamma}}{S^*_{111}}=|J_1|^{\Delta_\alpha}|J_2|^{\Delta_\beta}|J_3|^{\Delta_\gamma}\langle\phi_\alpha(-1)\phi_\beta(i)\phi_\gamma(0)\rangle_{\rm pl},
\end{align}
where $J_i$ is the Jacobian of the conformal mapping \eqref{z_to_w}. The initial states are
\begin{align}
|\phi^1\rangle&=\left(\frac{2\pi}{L}\right)^{-\Delta_\alpha}\lim_{z\rightarrow\infty}e^{2\pi z\Delta_\alpha/L}\phi_\alpha(z)|I^{\rm cyl}\rangle,\nonumber\\
|\phi^2\rangle&=\left(\frac{2\pi}{L}\right)^{-\Delta_\beta}\lim_{z\rightarrow i\infty}e^{-i2\pi z\Delta_\beta/L}\phi_\beta(z)|I^{\rm cyl}\rangle,\nonumber\\
|\phi^3\rangle&=\left(\frac{\sqrt{2}\pi}{L}\right)^{-\Delta_\gamma}\\
&\lim_{z\rightarrow{(-i-1)}\infty}e^{\frac{(i-1)}{\sqrt{2}}\frac{2\pi}{\sqrt{2}L}z\Delta_\gamma}\phi_\gamma(z)|I^{\rm cyl}\rangle\nonumber.
\end{align}
The Jacobian can be computed as 
\begin{align}
    |J_1| &= \left|\left(\frac{2\pi}{L}\right)^{-1}\lim_{z\rightarrow\infty}e^{2\pi z/L}w'(z)\right|\nonumber\\
    &=\left|\left(\frac{2\pi}{L}\right)^{-1}\lim_{w\rightarrow i}e^{2\pi z/L}\left(\frac{dz}{dw}\right)^{-1}\right|.
    \label{J1}
\end{align}
Using Eq.~\eqref{confmap_S}, the first and second term is 
\begin{align}
    e^{2\pi z/L} &= \exp[\ln\frac{w}{w-i}+i\ln\frac{w}{w+1}],\\
    \frac{dz}{dw} &= \frac{L}{2\pi}\left[-\frac{1}{w-i}-\frac{i}{w+1}+\frac{(1+i)}{w}\right].
\end{align}
Substituting these into Eq.~\eqref{J1},
\begin{align}
    |J_1| &= \bigg|\lim_{w\rightarrow i}\frac{w}{w-i}\exp\left[i\ln\frac{w}{w+1}\right]\nonumber\\
    &\quad \left(\left[-\frac{1}{w-i}-\frac{i}{w+1}+\frac{(1+i)}{w}\right]\right)^{-1}\bigg|\nonumber\\
    &=\left|\exp\left(i\ln\frac{i}{1+i}\right)\right|
    =e^{-\pi/4}.
\end{align}
In the same way, we can show $|J_2|=|J_3|=e^{-\pi/4}$. Thus, the 3-leg tensor is 
\begin{align}
    S^*_{\alpha\beta\gamma}
    &= e^{-\frac{\pi}{4}(\Delta_\alpha+\Delta_\beta+\Delta_\gamma)}\langle\phi_\alpha(-1)\phi_\beta(i)\phi_\gamma(0)\rangle_{\rm pl}.
\end{align}

\subsection{Conformal mapping of T}
The conformal mapping from the four-sided thermofield double state is 
\begin{align}
    z &= \frac{L}{2\pi}[-\ln(w-i)+\log(w+i)-i\ln(w+1)+i\ln(w-1)]\nonumber\\
    &=\frac{L}{2\pi}\left[\ln\left(\frac{w+i}{w-i}\right)+i\ln\left(\frac{w-1}{w+1}\right)\right].
\end{align}
To compute the Jacobian, we compute 
\begin{align}
    e^{2\pi z/L} &= \exp\left[\ln\frac{w+i}{w-i}+i\ln\frac{w-1}{w+1}\right],\\
    \frac{dz}{dw} &= \frac{L}{2\pi}\left[-\frac{1}{w-i}+\frac{1}{w+i}-\frac{i}{w+1}+\frac{i}{w-1}\right].
\end{align}
The Jacobian is then computed similarly as before:
\begin{align}
    |J_1|^{-1} &= \lim_{w\rightarrow i}\left|e^{-2\pi z/L}\left[-\frac{1}{w-i}+\frac{1}{w+i}-\frac{i}{w+1}+\frac{i}{w-1}\right]\right|\nonumber\\
    &=\frac{e^{\pi/2}}{2}.
\end{align}
The four-point function thus transforms as
\begin{align}
    \frac{T^*_{\alpha\beta\gamma\delta}}{T^*_{1111}}&=|J_1|^{\Delta_\alpha}|J_2|^{\Delta_\beta}|J_3|^{\Delta_\gamma}|J_4|^{\Delta_\delta}\langle\phi_\alpha(-1)\phi_\beta(i)\phi_\gamma(1)\phi_\delta(-i)\rangle_{\rm pl}\nonumber, \\
    &= \left(\frac{e^{\frac{\pi}{2}}}{2}\right)^{-\Delta_{\rm tot}}\langle\phi_\alpha(-1)\phi_\beta(i)\phi_\gamma(1)\phi_\delta(-i)\rangle_{\rm pl}.
\end{align}

\begin{figure}[tb]
    \centering
    \includegraphics[width=80mm]{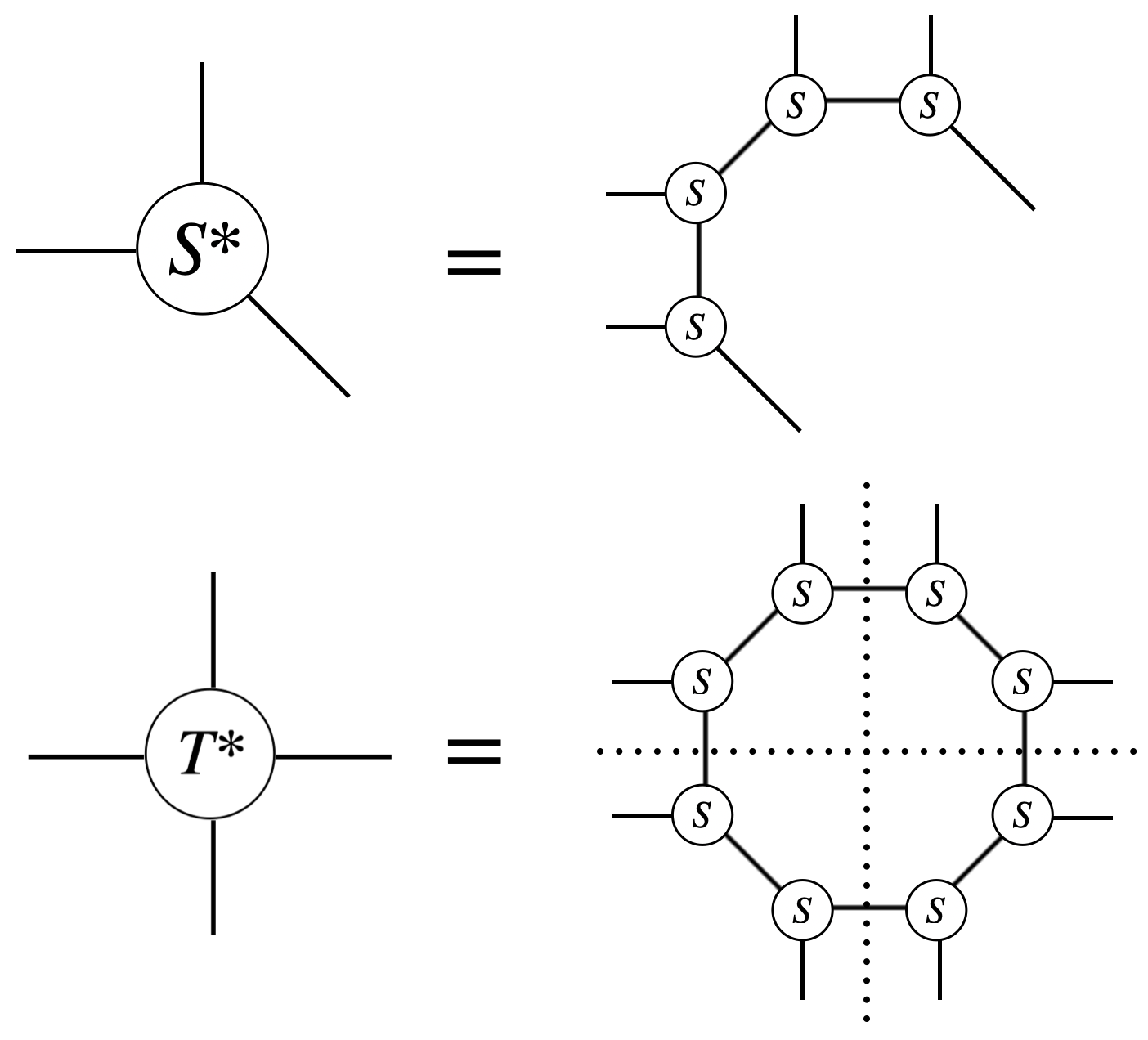}
    \caption{The contraction of the fixed-point tensors. We obtain $S$ from TRG and combine together to make $S^*$ and $T^*$. In this way, $T^*$ respects reflection symmetry along the dotted lines in addition to $C_4$ rotation symmetry. }
    \label{D4_TRG}
\end{figure}

\subsection{\texorpdfstring{${\rm D_4}$-symmetric TRG}{D4-symmetric TRG}}

We use the TRG scheme which aligns closely with the original paper's methodology~\cite{PhysRevLett.99.120601}. In principle, singular-value decomposition (SVD) of the four-leg tensor should yield two identical symmetric tensors, given the ${\rm D_4}$ symmetry of the original tensor. However, numerical errors sometimes make these two tensors non-identical. To mitigate this, we consistently select one of the three-leg tensors and supplement the other with its reflection. By adopting this approach, the fixed-point tensors, depicted in Fig.~\ref{D4_TRG}, maintain the ${\rm D_4}$ symmetry at every RG step by construction.
\subsection{Four-point function of the critical Ising model}
Here, we list the four-point function of the Ising model. Given the four coordinates $z_i$ and its cross-ratio $x\equiv (z_{12}z_{34})/({z_{13}z_{24}})$, the four-point functions of the Ising CFT are
\begin{align}
\langle \epsilon^4\rangle &= \left|\left[\prod_{1\leq i<j\leq 4} z_{ij}^{-\frac13}\right] \frac{1-x+x^2}{x^\frac23(1-x)^\frac23}\right|^2\nonumber,\\
\langle \sigma^2\epsilon^2\rangle &=\left|\left[z_{12}^\frac14 z_{34}^{-\frac58}\left(z_{13}z_{24}z_{14}z_{23}\right)^{-\frac{3}{16}} \right]\frac{1-\frac{x}{2}}{x^\frac38(1-x)^\frac{5}{16}}\right|^2\nonumber,\\
\langle \sigma^4\rangle &= |z_{13}z_{24}|^{-1/4} 
\frac{|1+\sqrt{1-x}|+|1-\sqrt{1-x}|}{2|x|^\frac14 |1-x|^\frac14}\nonumber.
\end{align}
The functions above are used to evaluate the analytic FP tensor elements in the main text.
\begin{figure}[bt]
    \centering
    \includegraphics[width=86mm]{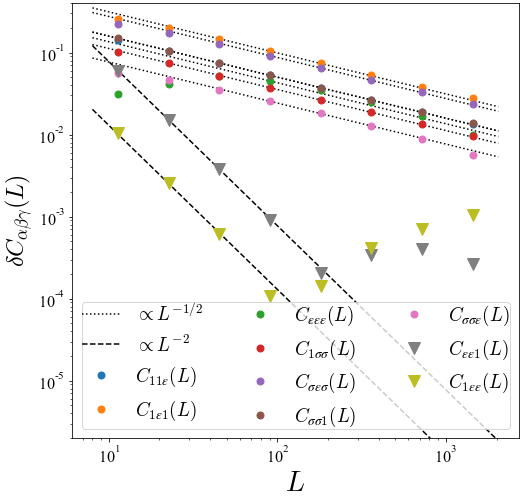}
    \caption{The finite-size corrections $\delta C_{\alpha\beta\gamma}(L)$ obtained from the numerical simulation of the critical Ising model. The numerical results for higher energy levels $\delta C_{\epsilon\epsilon 1}(L)$ and $\delta C_{1\epsilon\epsilon}(L)$ suffer from finite-$D$ effects for $L > 100$. The scalings of the finite-size corrections are nevertheless universal, which is consistent with Table III in Ref.~\cite{PhysRevB.107.155124}}
    \label{finite-size_ope}
\end{figure}

\section{Universal finite-size corrections}
Here, we discuss the finite-size corrections to Eq.~\eqref{ope_fomula}. The finite-size corrections of the OPE coefficients are defined as 
\begin{align}
    \delta C_{\alpha\beta\gamma}(L) = |C_{\alpha\beta\gamma} - C_{\alpha\beta\gamma}(L)|,
\end{align}
where $C_{\alpha\beta\gamma}(L)$ is defined in Eq.~\eqref{ope_fomula}. We found that $\delta C_{\alpha\beta\gamma}(L)$ exhibits a universal power-law decay as
\begin{align}
    \delta C_{\alpha\beta\gamma}(L) \sim L^{-p_{\alpha\beta\gamma}}.\label{our_p}
\end{align}
Our numerical results suggest $p_{\alpha\beta\gamma} =1/2$ for $({\alpha,\beta,\gamma}) = (1,1,\epsilon),$ $(1,\epsilon,1),$ $(\epsilon,\epsilon,\epsilon),$ $(1,\sigma,\sigma),$ $(\sigma,\epsilon,\sigma),$ $(\sigma,\sigma,1),$ and $(\sigma,\sigma,\epsilon)$, and $p_{\alpha\beta\gamma} =2$ for  $({\alpha,\beta,\gamma}) = (\epsilon,\epsilon,1)$ and $(1,\epsilon,\epsilon)$ as shown in Fig.~\ref{finite-size_ope}. Similar universal scalings were discussed in Ref.~\cite{PhysRevB.107.155124}, where they considered the overlap of critical wavefunctions $A_{\alpha\beta\gamma}=\langle \phi^{3*}_\gamma|\phi^1_\alpha\phi^2_\beta\rangle$. The three wavefunctions are defined on a ring with a circumference of $L_1$, $L_2$, and $L_3=L_1+L_2$, respectively, and the lower indices are the label of the corresponding primary states. Ref.~\cite{PhysRevB.107.155124} found the overlap of wavefunctions to be
\begin{align}
\frac{A_{\alpha\beta\gamma}}{A_{111}} & \sim \left[\left(\frac{L_3}{L_1}\right)^{\frac{L_1}{L_3}}\left(\frac{L_3}{L_2}\right)^{\frac{L_2}{L_3}}\right]^{-\frac{L_3}{L_1}\Delta_\alpha-\frac{L_3}{L_2}\Delta_\beta+\Delta_\gamma}C_{\alpha\beta\gamma}\nonumber\\
&+ \tilde{A}^{(p)}_{\alpha\beta\gamma}L_3^{-p_{\alpha\beta\gamma}},\label{previous_p}
\end{align}
where $p_{\alpha\beta\gamma}$ is the leading finite-size correction and $\tilde{A}^{(p)}_{\alpha\beta\gamma}$ is a prefactor that is independent of $L_3$. 

Our scaling exponents $p_{\alpha\beta\gamma}$ in Eq.~\eqref{our_p} coincide with those from the previous work in Eq.~\eqref{previous_p} for all fusion channels (see Table III of Ref.~\cite{PhysRevB.107.155124}). This universal scaling can be explained by considering rings 1 and 2 as an orbifold theory. The scaling $p_{\alpha\beta\gamma} =1/2$ is then attributed to the difference in the scaling dimensions of the orbifold theory, which is $\Delta_\epsilon/2=1/2$. (See Ref.~\cite{PhysRevB.107.155124} for details.) 
Similarly, we conjecture that the universal scaling for $\delta T_{\alpha\beta\gamma\delta} \sim L^{-1/3}$ can be understood by considering the three of four legs to be an orbifold theory.
\end{document}